\documentclass[%
 reprint,
 showpacs,
 aps,
 superscriptaddress,
 pre,]{revtex4-2}

\usepackage{hyperref} 
\usepackage{bm}
\usepackage{amsmath}
\usepackage{amssymb}
\usepackage{amsfonts}
\usepackage{braket}
\usepackage[dvipdfmx]{graphicx}
\usepackage{color}
\usepackage{here}
\usepackage{enumitem}
\usepackage{array}
\usepackage{titlesec}
\usepackage{braket}
\usepackage{booktabs}
\usepackage{comment}

\titleformat{\subparagraph}[runin]
  {\normalfont\normalsize\bfseries}{\thesubparagraph}{1em}{}
\titlespacing*{\subparagraph}
  {0pt}{1ex plus 0.5ex minus 0.5ex}{1em}

\begin{document}

\title{Simplification of Tensor Updates Toward Performance-Complexity Balanced Quantum Computer Simulation}

\author{Koichi Yanagisawa}
\affiliation{%
Information Technology R\&D Center, Mitsubishi Electric Corporation, Kanagawa 247-8501, Japan
}%
\author{Tsuyoshi Okubo}
\affiliation{%
Institute for Physics of Intelligence, The University of Tokyo, Bunkyo-ku, Tokyo 113-0033, Japan
}%
\author{Shota Koshikawa}
\affiliation{%
Information Technology R\&D Center, Mitsubishi Electric Corporation, Kanagawa 247-8501, Japan
}%
\author{Tsuyoshi Yoshida}
\affiliation{%
Information Technology R\&D Center, Mitsubishi Electric Corporation, Kanagawa 247-8501, Japan
}%
\author{Aruto Hosaka}
\affiliation{%
Information Technology R\&D Center, Mitsubishi Electric Corporation, Kanagawa 247-8501, Japan
}%
\author{Synge Todo}
\affiliation{%
Department of Physics, University of Tokyo, Tokyo 113-0033, Japan
}%
\affiliation{%
Institute for Physics of Intelligence, The University of Tokyo, Bunkyo-ku, Tokyo 113-0033, Japan
}%
\affiliation{%
Institute for Solid State Physics, The University of Tokyo, Kashiwa, Chiba 277-8581, Japan
}%


\begin{abstract}

Matrix Product States (MPS) provide a powerful framework for simulating quantum circuits. 
In practical simulations, tensor updates are typically performed in the canonical form (CF), which corresponds to the Schmidt decomposition and improves approximation accuracy. 
However, maintaining the canonical form introduces significant computational overhead. 
An alternative approach, known as the Simple Update (SU), does not enforce the Schmidt decomposition and is expected to reduce computational complexity. 
In this work, we systematically compare the performance and computational cost of SU and CF in quantum circuit simulations. 
We benchmark both methods on highly entangled circuits and on a QASM benchmark suite covering a wide range of circuit types. 
Our results show that SU achieves accuracy comparable to CF while reducing computational complexity, indicating that SU provides an efficient alternative for practical quantum circuit simulations.

\end{abstract}

\maketitle

\section{INTRODUCTION} \label{sec:introduction}

Quantum computing research is now entering the era of demonstrating early-stage fault tolerance, requiring many qubits \cite{logical_quantum_processor}. In this context, large-scale quantum computer simulation using classical computers has become more important than ever for finding potential applications or examining quantum systems without the need to construct and operate actual machines. Herein, the computational complexity is the bottleneck, which exponentially increases with the number of qubits \cite{quantum_circuit_simulator_40_qubits}.

The Tensor Network (TN) method, developed for solving many-body problems \cite{mixed_canonical_form, DMRG, DMRG_and_MPS,MERA,MPS_to_TPS,MPS_to_PEPS}, has gained attention in various large-scale problems.
For example, some aspects of TN technology are utilized in quantum computer simulations \cite{defeat_sycamore}.
Matrix Product State (MPS) is one of the best known formats of tensor networks \cite{tensor_networks_in_a_nutshell, mixed_canonical_form}.
State Vector (SV) describing a quantum state can be decomposed into MPS format approximately, and shallow circuits can be simulated with polynomial cost \cite{mixed_canonical_form, tensor_networks_in_a_nutshell, practical_introductiojn_to_tensor_networks, matrix_product_states_and_projected_entangled_pair_states}.

As quantum gates are applied and the state changes, the MPS tensors are updated.
For updating tensors, Canonical Form (CF) of MPS is helpful for improving the approximation accuracy \cite{mixed_canonical_form,canonical_form_user1}.
CF is useful for quantum circuits with neighboring quantum gates.
On the other hand, general quantum circuits have distant quantum gates, which refer to quantum gates that far separated from each other in a time series.
Distant quantum gates in quantum circuits make the simulation inefficient due to complexity proportional to distance to create CF.

Time-Evolving Block Decimation (TEBD) is an efficient simulation method proposed for 1D quantum many-body systems \cite{TEBD}.
Simple Update (SU), which utilizes tensors neighboring quantum gates in tensor update procedures \cite{simple_update}, is utilized in TEBD.
However, the systematic comparison between SU and CF is insufficient, and it remains unclear what circumstances SU can be effectively utilized.

To simulate an application of distant two-qubit gates, iterative operations of the QR decomposition in CF-based quantum computer simulations, leading to high complexity.
On the other hand, no iterative operations are needed in SU, leading to a reduction in complexity.
This work evaluates performance and complexity of SU for quantum computer simulations.
SU potentially shows a good performance-complexity tradeoff, for simplifying tensor updates compared with CF-based quantum computer simulations. 

The rest of the paper is organized as follows:
Sec.~\ref{sec:conventional_method} summarizes the TN methods with tensor updates based on CF and SU method.
Those are numerically evaluated in Sec.~\ref{sec:experiments}.
In Sec.~\ref{subsec:test1}, we employ quantum circuits for which SU is expected to be effective and complexity of SU is confirmed.
In Sec.~\ref{subsec:test2}, we employ various quantum circuits and performance of SU is evaluated.
In the end, Sec.~\ref{sec:conclusion} concludes the work.

\section{METHODS} \label{sec:conventional_method}

In this section, we review the TN methods for the quantum computer simulation.
First, we will start with the approximation of SV using MPS.
Second, we will move to the methods for tensor updates in MPS when a quantum gate operates on a quantum state.
Third, CF will be introduced and discussed the relation between CF and orthogonalized state.
Fourth, we discuss the problem in applying CF to the quantum computer simulation.
Finally SU is introduced.

\subsection{Approximation with MPS}

The state of $N$ qubits in quantum computers can be described with the SV as 
\begin{equation} \label{eq:SV}
|\psi\rangle = \sum_{\sigma_1 \sigma_2 \ldots \sigma_N} C^{\sigma_1 \sigma_2 \ldots \sigma_N} |{\sigma_1}\rangle |{\sigma_2}\rangle \ldots |{\sigma_N}\rangle,
\end{equation}
where $|\psi\rangle$, $C^{\sigma_1 \sigma_2 \ldots \sigma_N}$, and $\sigma_i \in \{0,1\}$ denote the state of $N$ qubits in quantum computers, SV, and the physical index for the state of $i$-th qubit \cite{tensor_networks_in_a_nutshell}, respectively.
It is difficult to deal with the SV, $C^{\sigma_1 \sigma_2 \ldots \sigma_N}$, exactly due to the numerous $(2^N)$ elements in the tensor.

\begin{figure}[t]
  \begin{center}
    \includegraphics[clip, width=8.4cm]{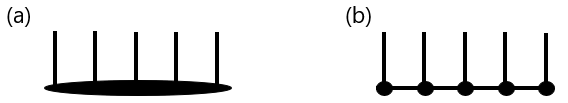}
    \caption{Tensor network diagrams of (a) SV and (b) MPS, where the nodes and the lines represent tensors and indices, respectively. The connected lines indicate a contraction. In this example, SV is represented by a rank-5 tensor, while the MPS consists of three rank-3 tensors and two rank-2 tensors.}
    \label{fig:network_topology_variation}
  \end{center}
\end{figure}

To reduce the number of independent elements, we employ MPS as shown in FIG.~\ref{fig:network_topology_variation} to approximate the rank-$N$ tensor $C^{\sigma_1 \sigma_2 \ldots \sigma_N}$ using a set of rank-3 local tensors $\{ M^{\sigma_i} \}$.
The approximation of SV by MPS is expressed as 
\begin{equation} \label{eq:MPS}
C^{\sigma_1 \sigma_2 \ldots \sigma_N} \approx M^{\sigma_1} M^{\sigma_2} \ldots M^{\sigma_N},
\end{equation}
where $M^{\sigma_i}$ denotes a rank-3 tensor. 
We also express the local tensor $M^{\sigma_i}$ with virtual indices $a_i,a_{i+1}$ explicitly as $M^{\sigma_i}_{a_i,a_{i+1}}$.
We apply this notation to other tensors as well.
The tensor multiplication follows a contraction rule, 
\begin{align} \label{eq:contraction_rule}
\{ M^{\sigma_{i-1}} M^{\sigma_{i}} \}_{a_{i-1}, a_{i+1}} = \sum_{a_{i}=1}^D M_{a_{i-1},a_{i}}^{\sigma_{i-1}} M_{a_{i},a_{i+1}}^{\sigma_{i}}, 
\end{align}
where $D$ denotes a degree of freedom for index $a_i$, called bond-dimension.

Each local tensor $M^{\sigma_i}$ has $2D^2$ elements, which is assumed not to depend on $N$, enabling the simulation of quantum computers with a reduced number of elements.
Unless otherwise stated, we will employ MPS to describe quantum states throughout this paper.

\subsection{Tensor updates in MPS} \label{sec:tensors_updates_in_MPS}

In the MPS-based simulation, applying quantum gates is simulated through tensor updates in MPS.
In detail, quantum gates are described with tensors, and applied through tensor contraction with MPS \cite{tensor_networks_in_a_nutshell}, resulting in a new MPS.
Quantum gates can be decomposed into the elemental gates, including one-qubit gates and two-qubit gates \cite{universality_of_quantum_gates}. 
Here, we explain the processes of applying one-qubit gates, two-qubit gates between adjacent qubits, and long-range two-qubit gates, respectively.

\subparagraph{One-qubit gate}

A one-qubit gate operation is described with the contraction of a local tensor $M_{a_i,a_i+1}^{\sigma_{i}}$ with a matrix $G^{\sigma_{i}' \sigma_{i}}$ representing the quantum gate, i.e.,
\begin{align} \label{eq:applying_a_quantum_gate_for_one_qubit}
  M_{a_{i-1},a_i}^{\sigma_i} \rightarrow {M'}_{a_{i-1},a_i}^{\sigma_{i}} = \sum_{\sigma_{i}'} M_{a_{i-1},a_i}^{\sigma_{i}'} G^{\sigma_{i}' \sigma_{i}}.
\end{align}
After this one-qubit gate operation, ${M'}_{a_i,a_i+1}^{\sigma_{i}}$ becomes the new local tensor.
/
\subparagraph{Two-qubit gate between adjacent qubits}
\label{subsec:adjacent_two_qubit_gate}

A two-qubit gate is expressed by a rank-4 tensor.
An two-qubit gate operation between adjacent qubits is denoted as
\begin{align} \label{eq:applying_a_quantum_gate_for_two_adjacent_qubits}
  &\sum_{a_{i}=1}^D M_{a_{i-1},a_{i}}^{\sigma_i} M_{a_{i},a_{i+1}}^{\sigma_{i+1}} \nonumber \\
  &\rightarrow  \sum_{\sigma_{i}', \sigma_{i+1}', a_{i}} M_{a_{i-1},a_{i}}^{\sigma_{i}'} M_{a_{i},a_{i+1}}^{\sigma_{i+1}'} G^{\sigma_{i}' \sigma_{i} \sigma_{i+1}' \sigma_{i+1}}
  = T_{a_{i-1},a_{i+1}}^{\sigma_{i} \sigma_{i+1}},
\end{align}
where $G^{\sigma_{i}' \sigma_{i} \sigma_{i+1}' \sigma_{i+1}} $ is a rank-4 tensor representing the two-qubit gate between adjacent qubits, and $T_{a_{i-1},a_{i+1}}^{\sigma_{i} \sigma_{i+1}}$ is the rank-4 tensor contracted over $\sigma_{i}'$, $\sigma_{i+1}'$, and $a_{i}$.
After the gate operation, maintaining the network topology is needed.
This process includes following steps:
 
 \noindent STEP1: Reshape the tensor $T_{a_{i-1},a_{i+1}}^{\sigma_i,\sigma_{i+1}}$ into a matrix $\tilde{T}_{(a_{i-1},\sigma_{i}),(a_{i+1},\sigma_{i+1})}$.

 \noindent STEP2: Decompose the matrix $\tilde{T}_{(a_{i-1},\sigma_{i}),(a_{i+1},\sigma_{i+1})}$ into $\sum_{k=1}^{D'} A_{a_{i-1},k}^{\sigma_{i}} S_{k,k} B_{k, a_{i+1}}^{\sigma_{i+1}}$ using SVD, where $A$ and $B$ are tensors reshaped from unitary matrices and $S$ is a diagonal matrix with singular values. 

 \noindent STEP3: Contract tensors between $A$ and $S$ or $B$ and $S$ to make two rank-3 tensors $\sum_{a_{i}}^{D'} M_{a_{i-1},a_{i}}^{\sigma_i} M_{a_{i},a_{i+1}}^{\sigma_{i+1}}$.

The degree of freedom is changed from $D$ to $D'=2D$ during the process.
Under certain conditions, e.g., $D'$ exceeds a threshold, $D'$ is truncated by the low-rank approximation based on SVD \cite{low_rank_approximation}.

\subparagraph{Long-range two-qubit gate}
\label{subsec:non_adjacent_two_qubit_gate}

In this subsection, we discuss the process of applying two-qubit gates between $i$-th and $j$-th qubit.

When $i$ and $j$ are not nearest neighbors, the method discussed in Sec. \ref{subsec:adjacent_two_qubit_gate} cannot be applied directly.
Swapping the physical indices, $\sigma_i$ and $\sigma_{i+1}$, as
\begin{align} \label{eq:swap_two_adjacent_tensors}
  &\sum_{a_{i}=1}^{D} M_{a_{i-1},a_{i}}^{\sigma_i} M_{a_{i},a_{i+1}}^{\sigma_{i+1}} \nonumber \\
  &\approx \sum_{a_{i}=1}^{D} \tilde{M}_{a_{i-1},a_{i}}^{\sigma_{i+1}} \tilde{M}_{a_{i},a_{i+1}}^{\sigma_{i}},
\end{align}
is one of methods to applying long-range two-qubit gates \cite{cuQuantum}.
Swap gate is one of qubit gates between qubits \cite{tensor_networks_in_a_nutshell} and we utilize the process explained in Sec.~\ref{subsec:adjacent_two_qubit_gate}.
The physical index $\sigma_i$ in Eq.~\eqref{eq:swap_two_adjacent_tensors} is moved repeatedly until being adjacent to the local tensors with $\sigma_j$.
After contracting the two-qubit gate, swapping physical indices is again repeatedly applied to match the index order before and after the operation.

\subsection{Canonical form}
\label{subsec:decomposition_method_in_this_paper}
The MPS format can be approximated by low-rank approximation, which truncates some vectors based on tensors decomposition.
CF is a description of decomposed tensors \cite{mixed_canonical_form} and influence the performance of the approximation.
The qubit state in Eq.~\eqref{eq:SV} is described with the Schmidt decomposition
\begin{align} \label{eq:Schmidt_decomposition}
        |\psi\rangle = \Sigma_k S_{k,k} \ket{x_k}_X \ket{y_k}_Y,
\end{align}
given by SVD of SV $C^{\sigma_1 \sigma_2 \ldots \sigma_N} = \Sigma_k U_{\sigma_1 \sigma_2 \ldots \sigma_m, k} S_{k,k} V^{*}_{\sigma_{m+1} \ldots \sigma_N,k}$ with the decomposition subsystems $X$ and $Y$, Schmidt bases $x_k = \sum_{\sigma_{1} \ldots \sigma_m} U_{\sigma_1 \sigma_2 \ldots \sigma_m, k} |{\sigma_1}\rangle |{\sigma_2}\rangle \ldots |{\sigma_m}\rangle $, and $ y_k = \sum_{\sigma_{m+1} \ldots \sigma_N} V^{*}_{\sigma_{m+1} \ldots \sigma_N,k} |{\sigma_{m+1}}\rangle \ldots |{\sigma_N}\rangle$, and Schmidt coefficients $S_{k,k}$ \cite{mixed_canonical_form}.
Here, singular values and unitary matrices in SVD corresponds to Schmidt coefficients and basis transformation, respectively.
Truncating the Schmidt coefficients except for ones having $D$ largest values and related Schmidt bases in Eq.~\eqref{eq:Schmidt_decomposition} can minimize the error with the low-rank approximation of SV under a given $D$.

Now, we discuss the Schmidt decomposition in Eq.~\eqref{eq:Schmidt_decomposition} of quantum states in the MPS format.
CF is useful form represented as
\begin{align} \label{eq:canonical_form}
  C^{\boldsymbol{\sigma}} = A^{\sigma_1} \ldots A^{\sigma_m} \Lambda B^{\sigma_{m+1}} \ldots B^N,
\end{align}
where $\boldsymbol{\sigma}$ denotes $\{\sigma_1 \sigma_2 \ldots \sigma_N\}$, $A^{\sigma_l} = A_{a_{l-1},a_l}^{\sigma_l}$ and $B^{\sigma_l} = B_{a_{l-1},a_l}^{\sigma_l}$ satisfy canonical conditions, in other words, $\sum_{a_{l-1}, \sigma_l} A_{a_{l-1}, a'_{l}}^{* \sigma_l} A_{a_{l-1}, a_l}^{\sigma_l}$ and $\sum_{a_l, \sigma_l} B_{a_{l-1}, a_l}^{\sigma_l} B_{a'_{l-1}, a_{l}}^{* \sigma_l}$ are equal to the identity matrix, $m$ denotes the position of the Schmidt coefficients, and $\Lambda$ denotes a canonical center \cite{mixed_canonical_form}.
Eq.~\eqref{eq:canonical_form} is also called mixed canonical form.
Matrix form is also utilized, e.g., $\sum_{\sigma_l} A^{\sigma_l \dag} A^{\sigma_l} = I$, where $\dag$ denotes Hermitian conjugate and $I$ is the identity matrix.
In CF, the local matrices with a physical index in subsystem X are transformed into left normalized matrices $\{A^{\sigma_l}\}$ and those in subsystem Y are transformed into right normallized matrices $\{B^{\sigma_l}\}$.
$\left(A^{\sigma_1} \ldots A^{\sigma_{m-1}}\right)$ in Eq. \eqref{eq:canonical_form} is obtained from the iterative operations of the QR decomposition $M_{(\sigma_i, a_{l-1}), a_{l}} = \sum_{a'_{l}} Q_{(\sigma_i, a_{l-1}), a'_{l}} R^{i}_{a'_{l}, a_{l}} (i = 1, \cdots, m-1)$ and the contraction $R^{i} M^{\sigma_{i+1}} = \tilde{M}^{\sigma_{i+1}} (i = 1, \cdots, m-2)$, where $M_{(\sigma_i, a_{l-1}), a_{l}}$ is the matrix reshaped from $M^{\sigma_i}$, $Q^{i}$ and $R^{i}$ are the matrices given by the QR decomposition from $M^{\sigma_i}$, $A^{\sigma_i}$ is the tensor reshaped from $Q^i$, and a tensor with $\~$ denotes the updated tensor in a operation.
$\~$ is abbreviated in the next operation.
$\left(\tilde{B}^{\sigma_{m+1}} B^{\sigma_{m+2}} \ldots B^N\right)$ is also obtained from the iterative operations of the QR decomposition $M_{a_{l-1}, (\sigma_i, a_{l})}= \sum_{a'_{l-1}} R^{i}_{a_{l-1}, a'_{l-1}} Q^{i}_{a'_{l-1}, (\sigma_i, a_{l})} (i = N, N-1, \cdots, m+1)$ and the contraction $M^{\sigma_{i-1}} R^{i} = \tilde{M}^{\sigma_{i-1}} (i=N,N-1,\cdots,m+2)$, where $M_{a_{l-1}, (\sigma_i, a_{l})}$ is the matrix reshaped from $M^{\sigma_i}$, $Q^{i}$ and $R^{i}$ are the matrices given by the QR decomposition from $M^{\sigma_i}$, and $B^{\sigma_i}$ is the tensor reshaped from $Q^i$.
$A^{\sigma_m}$ and $\Lambda$ are obtained from SVD of $\tilde{M}^{\sigma_m} = R^{m-1} M^{\sigma_m} R^{m+1}$, i.e., $\tilde{M}_{(\sigma_m, a_{m-1}), a_{m} } = \sum_{a'_{m}} U_{(\sigma_m, a_{m-1}), a'_{m}} \Lambda_{a'_{m}, a'_{m}} V^{*}_{a_{m}, a'_{m}}$, where $\tilde{M}_{(\sigma_m, a_{m-1}), a_{m}}$ is the matrix reshaped from $\tilde{M}^{\sigma_m}$, $U$ and $V$ are the unitary matrices and $\Lambda$ is the diagonal matrix given by the SVD decomposition from $\tilde{M}^{\sigma_m}$, and $A^{\sigma_m}$ is the tensor reshaped from $U$. 
$V^{\dag}$ is contracted with $B^{\sigma_{m+1}}$.
CF of MPS corresponds to Schmidt decomposition \cite{mixed_canonical_form}, i.e., $\left(A^{\sigma_1} \ldots A^{\sigma_m}\right)$, $\left(B^{\sigma_{m+1}} \ldots B^N\right)$ and $\Lambda$ correspond to Schmidt bases ($x_k$, and $y_k$) and Schmidt coefficients $S_{k,k}$, respectively.
Note that, this paper employs S as singular values of arbitrary matrix regardless whether these are Schmidt coefficients, while $\Lambda$ as Schmidt coefficients.
In other words, singular values of SV or the canonical center are described with either $S$ or $\Lambda$, while singular values of other matrices are only denoted by $S$.
In CF-based quantum computer simulations, the truncation of the diagonal matrix $\Lambda$ in CF is equivalent to the truncation of $S$ in Eq.~\eqref{eq:Schmidt_decomposition} given by rank-$N$ tensor $C^{\sigma_1 \sigma_2 \ldots \sigma_N}$.

\subsection{Problems of CF-based quantum computer simulation}\label{subsec:quantum_gate_placement_and_computational_cost}

\begin{figure}[t]
  \begin{center}
    \includegraphics[clip, width=8.4cm]{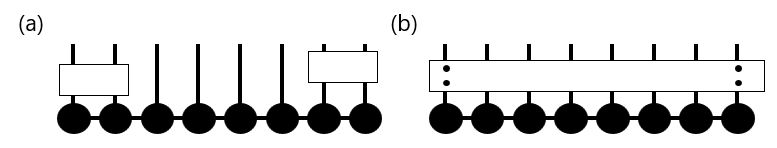}
    \caption{The difinition of terms: (a) distant gates, (b) a long-range gate.}
    \label{fig:explanation_of_terms}
  \end{center}
\end{figure}

In this subsection, we discuss problems of CF-based quantum computer simulation.
Quantum computings generally require two-qubit gates.
To simulate an application of distant two-qubit gates, the position of the canonical center $m$ in Eq.~\eqref{eq:canonical_form} should be changed, which needs iterative operations of the QR decomposition as discussed in Sec. \ref{subsec:decomposition_method_in_this_paper}.
Note that this paper employs the term \textit{distant gates} to refer to quantum gates which are far separated from each other in a time series, and the term \textit{a long-range gate} to refer to a single gate applied to far qubits as shown in FIG.~\ref{fig:explanation_of_terms}.
The complexity increases depending on the distance between quantum gates because the position of the canonical center change among the consecutive time step.
Thus, in some quantum circuits, CF leads high complexity due to iterative operations of the QR decomposition.

\subsection{Simple update} \label{subsec:simple_update}

In order to address the problems above, we evaluate simplified tensor update methods, which potentially shows efficient quantum computer simulation with distant two-qubit gates, and its performance-complexity tradeoff is numerically analyzed in Sec.~\ref{sec:experiments}.

\begin{figure}[t]
  \begin{center}
    \includegraphics[clip, width=8.4cm]{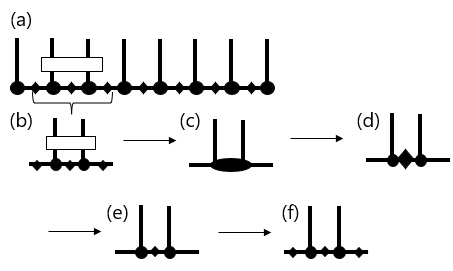}
    \caption{The procedure of SU: (a) ansatz of the SU, (b) tensors around quantum gate, (c) tensor after contraction, (d) decomposed tensors, (e) approximated tensors, and (f) reverted form of tensors. }
    \label{fig:simple_update_procedure}
  \end{center}
\end{figure}

For the tensor updates simplification, we focus on SU \cite{simple_update}, which is widely used and demonstrate high performance with low complexity for simulating quantum many-body systems \cite{SU_user_michael1, SU_user_michael2}.
SU was originally proposed for TEBD, which is efficient simulation for 1D quantum many-body systems \cite{TEBD}.
SU is also utilized for efficient quantum circuit simulations \cite{quantum_circuits_with_2D_tensor_network}, and parallelization methods are proposed \cite{parallelization_using_TEBD}.
In this paper, we quantitatively evaluate the performance of SU for quantum circuits with distant two-qubit gates.
We expect that SU is also suitable for these quantum circuits.
FIG.~\ref{fig:simple_update_procedure} shows the SU procedure.
Adjacent two-qubit gate operation with SU consists of the following steps:

 \noindent (a) Use Vidal canonical form \cite{TEBD} as an ansatz which matrices with Schmidt coefficients $S^j (j=1,\cdots,N-1)$ are set between local tensors having a physical index $M^{\sigma_1}S^1 M^{\sigma_2} \ldots S^{N-1} M^{\sigma_N}$ with the two-qubit gate $G^{\sigma_{i}' \sigma_{i} \sigma_{i+1}' \sigma_{i+1}}$.

 \noindent (b) Select tensors around two-qubit gate from MPS as $\sum_{\bm{\sigma}',\bm{k}} S_{a_{i-1},k_{i-1}}^{i-1} M_{k_{i-1},k_{i}}^{\sigma_{i}'} S_{k_i,k_i}^i M_{k_i,k_{i+1}}^{\sigma_{i+1}'} S_{k_{i+1},a_{i+1}}^{i+1} G^{\sigma_{i}' \sigma_{i} \sigma_{i+1}' \sigma_{i+1}}$.

 \noindent (c) Contract selected tensors as $M_{a_{i-1},a_{i+1}}^{\sigma_i, \sigma_{i+1}}$.

 \noindent (d) Decompose the contracted tensors by SVD as $\sum_{a_{i}=1}^{D'} U_{a_{i-1},a_{i}}^{\sigma_i} S_{a_i,a_i} V_{a_{i},a_{i+1}}^{\dagger \sigma_{i+1}}$, where $D' = \mathrm{min}(2 \times D_{a_{i-1}},2 \times D_{a_{i+1}})$.

 \noindent (e) Truncate the diagonal matrix made by SVD as $\sum_{a_{i}=1}^{D} U_{a_{i-1},a_{i}}^{\sigma_i} S_{a_i,a_i} V_{a_{i},a_{i+1}}^{\dagger \sigma_{i+1}}$, where $D = \mathrm{min}(2 \times D_{a_{i-1}},2 \times D_{a_{i+1}}, D_{\mathrm{max}})$ and $D_{\mathrm{max}}$ is a hyperparameter.

 \noindent (f) Revert contracted adjacent diagonal tensors as $\sum S_{a_{i-1},a_{i-1}}^{i-1} M_{a_{i-1},a_{i}}^{' \sigma_{i}} S_{a_i,a_i}^{' i} M_{a_i,a_{i+1}}^{' \sigma_{i+1}} S_{a_{i+1},a_{i+1}}^{i+1}$, where $S_{a_{i-1},a_{i-1}}^{i-1} M_{a_{i-1},a_{i}}^{' \sigma_{i}} = U_{a_{i-1},a_{i}}^{\sigma_i}$ and $M_{a_{i},a_{i+1}}^{' \sigma_{i+1}} S_{a_{i+1},a_{i+1}}^{i+1} = V_{a_{i},a_{i+1}}^{\dagger \sigma_{i+1}}$.

 \noindent A two-qubit gate operation between distant qubits is executed with iteration of swap gates, and the process of the SU also includes the steps explained above. 
Thus, SU utilizes only adjacent singular values in the procedure and can reduce the complexity compared with deriving CF.

Although SU utilizes only adjacent singular values in the procedure, SU keeps canonical conditions under the following conditions \cite{TEBD_and_error}:

 \noindent (i) $A^i = M^{\sigma_i} S^i, B^i = S^{i-1} M^i for \forall i \in \{1, \cdots, N \}$ in the initial condition satisfies canonical conditions.

 \noindent (ii) The applied quantum gate is unitary.

 \noindent (iii) The diagonal matrix is not truncated in process (e) of FIG.~\ref{fig:simple_update_procedure}.

Let us consider whether above conditions are satisfied in the quantum circuit simulation.
First, the initial condition starts from $ M^{\sigma_i = 0} = \begin{bmatrix}1\end{bmatrix}, M^{\sigma_i = 1} = \begin{bmatrix}0\end{bmatrix}, S^{i} = \begin{bmatrix}1\end{bmatrix} for \forall i \in \{1, \cdots, N \}$ to express $|00....0\rangle$ and (i) is satisfied.
Second, Unitary quantum gates are applied and (ii) is satisfied.
However, the diagonal matrix is truncated and (iii) is not satisfied.
Therefore, the performance of SU could decrease due to truncation, depending on quantum circuits.
We will investigate the performance of SU quantitatively in representative example circuits by numerical simulation in Sec.~\ref{sec:experiments}.

\section{EVALUATION} \label{sec:experiments}

In Sec.~\ref{subsec:test1}, we employ a shallow quantum circuit with distant two-qubit gates between adjacent qubits.
SU is expected to be effective for this quantum circuit simulation and complexity of SU is confirmed.
In Sec.~\ref{subsec:test2}, we employ various quantum circuits, and performance of SU is investigated.

\subsection{Confirming Reduction in Complexity}\label{subsec:test1}

\begin{figure}[t]
  \begin{center}
    \includegraphics[clip, width=8.4cm]{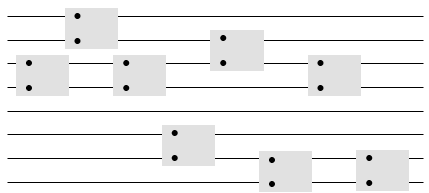}
    \caption{Exemplified shallow quantum circuit with distant two-qubit gates between adjacent qubits. Though this figure includes only 8 qubits for a small example, we evaluated up to 2000 qubits.}
    \label{fig:quantum_circuit_with_adjacent_gates.png}
  \end{center}
\end{figure}

In the first test case, we employ a shallow quantum circuit with distant two-qubit gates between adjacent qubits as shown in FIG.~\ref{fig:quantum_circuit_with_adjacent_gates.png}.
Since this quantum circuit is composed of distant two-qubit gates, the canonical center $m$ needs to be moved each time two-qubit gates are applied, as discussed in Sec.~\ref{subsec:quantum_gate_placement_and_computational_cost} on CF-based tensor updates.
On the other hand, SU skips the procedure.
Therefore, SU is expected to reduce the complexity due to distant two-qubit gates.
Furthermore, since this quantum circuit is shallow and the two-qubit gates are applied to adjacent qubits, small truncations are expected to be needed.
This fact indicates SU shows high performance in this circuit.

The two-qubit gates order is randomly chosen and the two-qubit gates allocation can be nonuniform.
In other words, some qubits can have more two-qubit gates operation than the others.
We execute multiple iterations by changing the seed of the pseudorandom number generator.
This evaluation excluded the optimization of the quantum gates order.
Two truncation parameters were employed: the number of singular values $\le 5$, and the singular values are truncated when the ratio relative to largest singular value is less than $0.0001$.
The number of two-qubit gates $N$ equals the number of qubits, where $N$ is a parameter.

In this setup, we quantify the computation runtime as a complexity metric and fidelity referenced to the quantum state given by CF-based quantum simulation as a performance metric;
\begin{align} \label{eq:fidelity_between_SU_and_CF}
F_{\mathrm{CF}} = |\langle\psi_{\mathrm{CF}}|\psi_{\mathrm{SU}}\rangle|^{2},
\end{align}
where $|\psi_{\mathrm{CF}}\rangle$ and $|\psi_{\mathrm{SU}}\rangle$ denotes the qubit states given by CF and SU, respectively.
In this paper, CF-based tensor update is abbreviated as CF.
Runtime and fidelity were quantified by computing the mean and standard error across four independently generated random circuits.

\begin{figure}[t]
  \begin{center}
    \includegraphics[clip, width=8.4cm]{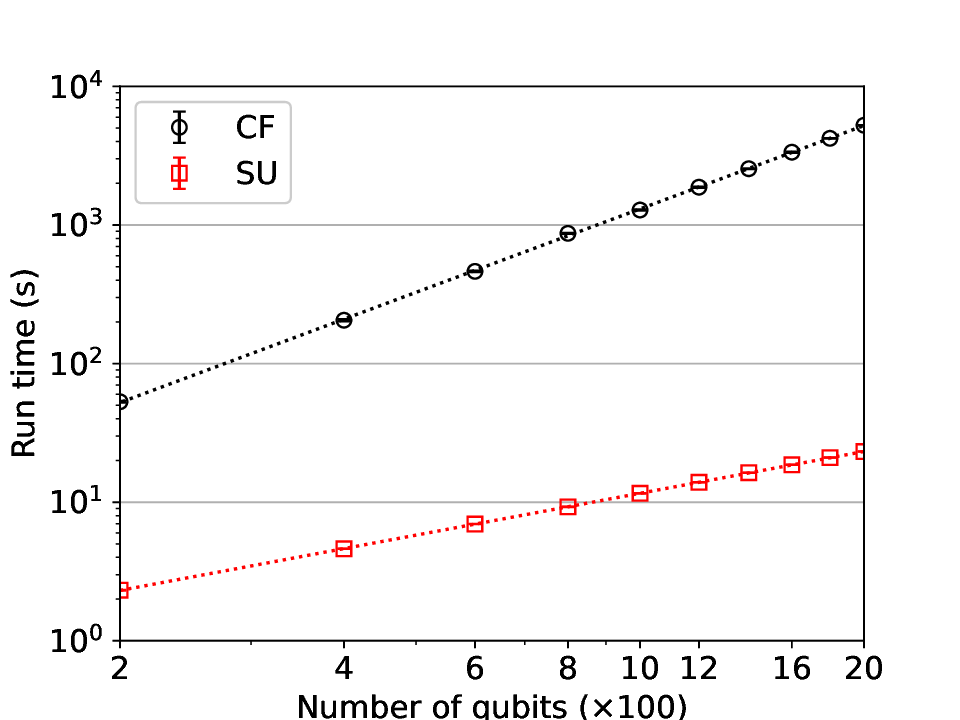}
    \caption{Runtime in tensor updates through numerical simulation.
    Dotted lines are regression lines.
    SU significantly reduces the runtime for a large number of qubits $N$.
    Log-scale slopes are around 2 for CF and 1 for SU, respectively.
    Note that CF refers to the CF-based tensor update.
    The error bars represent standard errors.
    }
    \label{fig:regression_analysis_for_run_time.eps}
  \end{center}
\end{figure}

FIG.~\ref{fig:regression_analysis_for_run_time.eps} shows the runtime for quantum circuits.
In detail, the runtime refers to the duration required to update MPS using given tensor-form representation of the quantum circuit, excluding the time required to compute physical observables from the MPS.
The SU significantly reduced the runtime in the cases of a large number of qubits $N$, e.g., SU is approximately $230$ times faster than CF for $N = 2000$.
In a log scale view, the slope of the runtime against $N$ is approximately 2 for CF and 1 for SU, respectively.
Roughly speaking, the runtime is proportional to $N^2$ for CF and proportional to $N$ for SU, thus SU reduces the computational costs by $O(N)$.
This reduction in computational costs is understood as follows.
The average distance between two time-consective two-qubit gates is $O(N)$ for randomly ordered two-qubit gates, which requires the change of position $m$ in Eq.\eqref{eq:canonical_form} at $O(N)$ complexity.
CF requires the $O(N)$ complexity given by the QR decomposition and absorbing the matrix $R$, but SU does not.
\begin{figure}[t]
  \begin{center}
    \includegraphics[clip, width=8.4cm]{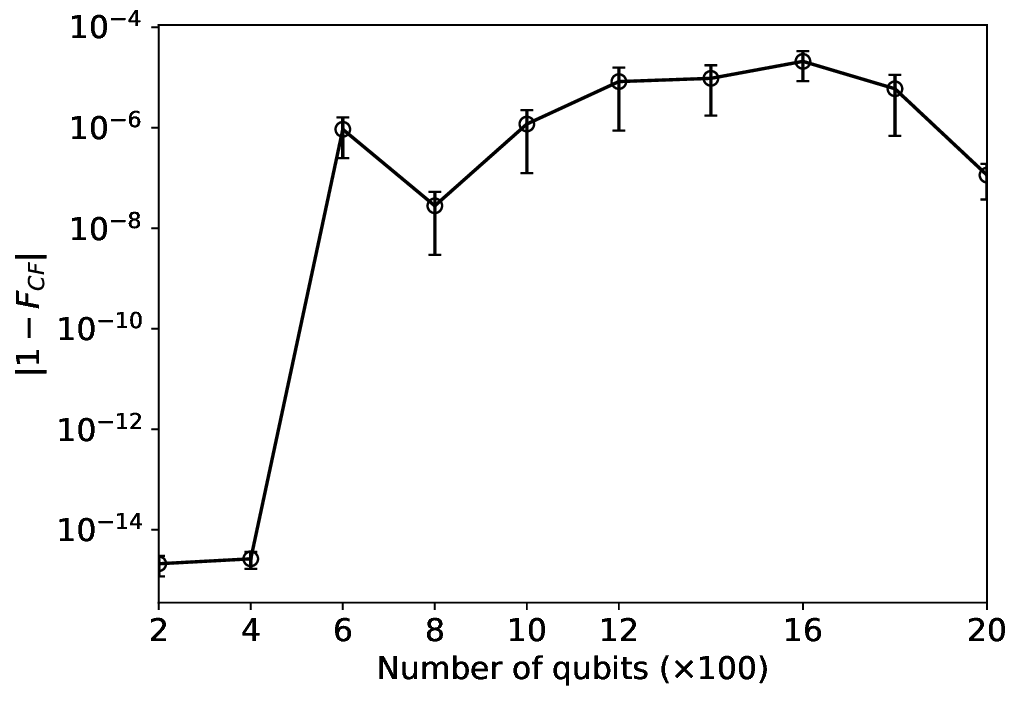}
    \caption{
	    Numerically obtained fidelity $F_{\mathrm{CF}}$ (defined in Eq.~\eqref{eq:fidelity_between_SU_and_CF}).
        The absolute values are calculated to avoid numerical errors near zero.
        Qubits states given by CF and SU are almost identical.
        Note that CF refers to the CF-based tensor update.
        The error bars represent standard errors.
        }
    \label{fig:data_with_adjacent_gates/compare_fidelities.eps}
  \end{center}
\end{figure}
FIG.~\ref{fig:data_with_adjacent_gates/compare_fidelities.eps} shows the fidelity $F_{\mathrm{CF}}$ in Eq.~\eqref{eq:fidelity_between_SU_and_CF}.
The fidelity is almost $1$, i.e., the qubit states given by CF and SU are almost the same.
According to these results, SU shows a good balance of performance (fidelity) and complexity (computational cost, i.e., runtime) in MPS-based quantum computer simulation.

\subsection{Investigating Performance}\label{subsec:test2}

The aim of the second test case is to investigate the performance in various quantum circuits.
In the previous subsection, the quantum circuit is shallow and is composed of two-qubit gates between adjacent qubits.
In this conditions, small truncations are expected, and SU showed high performance.
On the other hand, quantum state given by SU is considered to deviate from the quantum state given by CF under a highly entangled condition.
Magnitude of the error under highly entangled conditions is unknown. 

\begin{figure}[t]
  \begin{center}
    \includegraphics[clip, width=8.4cm]{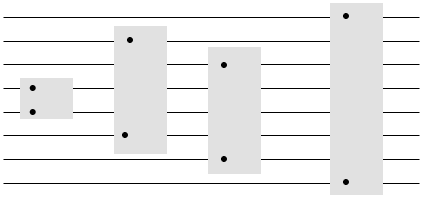}
    \caption{Exemplified quantum circuit for fidelity analysis. This kind of circuits is employed for the quantum volume measurement. Circles represent qubits for the quantum gate operation.}
    \label{fig:quantum_volume_circuit}
  \end{center}
\end{figure}

To investigate the performance under a highly entangled condition, we consider circuits with long-range two-qubit gates as shown in FIG.~\ref{fig:quantum_volume_circuit}.
This kind of circuits is employed to measure the quantum volume \cite{Quantum_Volume}.
For setting up the quantum circuit, $\mathit{Qiskit}$ library was employed \cite{qiskit2024}. FIG.~\ref{fig:quantum_volume_circuit} focuses on the area with the eight qubits and the depth one, where the depth refers to the layers of SU(4) operations in the quantum circuit \cite{qiskit_quantum_volume_circuit}.
In this test case, we employ the depth as variable and decompose quantum gates into elemental quantum gates operated over one or two qubits.
In this quantum circuit, a two-qubit gate is applied to a random pair of qubits.
The quantum state is likely to be highly entangled, which is expected to lead to many truncations.
Two truncation parameters were employed: the number of singular values $\le 10$, and the singular values are truncated when the ratio relative to largest singular value is less than $0.0001$.
The number of qubits $N$ is $15$.
We examine $29$ random quantum circuits generated by Qiskit for each depth.

\begin{figure}[t]
  \begin{center}
    \includegraphics[clip, width=8.4cm]{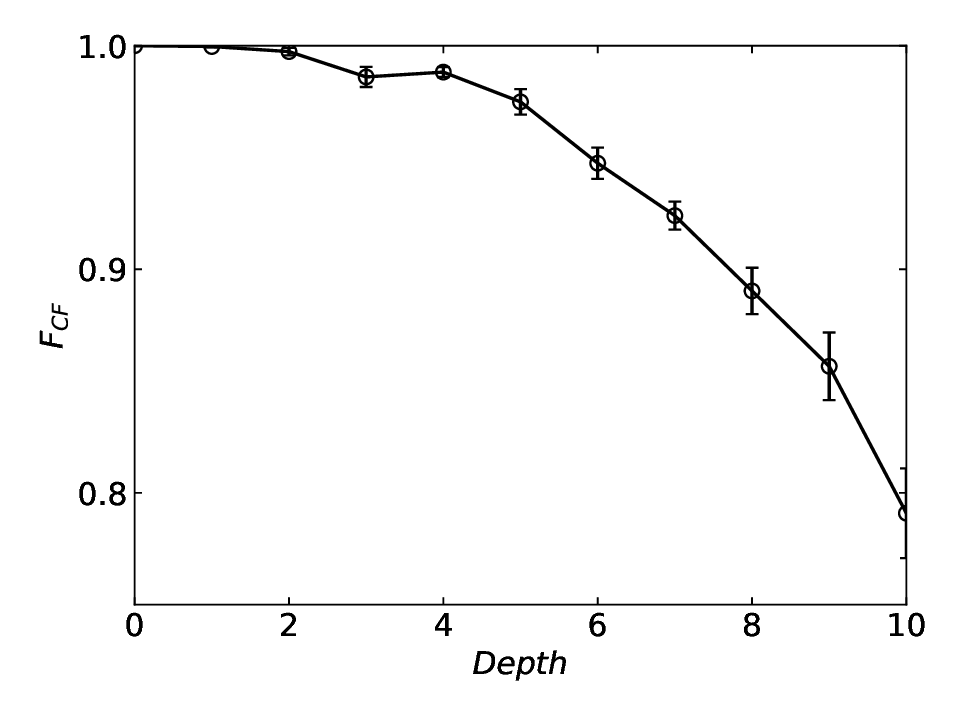}
    \caption{
             Numerically obtained fidelity $F_{\mathrm{CF}}$ (defined in Eq.~\eqref{eq:fidelity_between_SU_and_CF}).
             The fidelity was close to 1 under the shallow depth condition. 
             On the other hand, the fidelity decreased under the deep depth condition.
             The error bars represent standard errors.
             }
    \label{fig:compare_fidelities_between_SU_and_CF}
  \end{center}
\end{figure}

FIG.~\ref{fig:compare_fidelities_between_SU_and_CF} shows the average of the obtained fidelity of SU referenced to the quantum state given by CF (Eq.\eqref{eq:fidelity_between_SU_and_CF}).
The fidelity decreased under the deep depth condition.
This supports our hypothesis that quantum state given by SU largely deviates from quantum state given by CF under highly entangled conditions.

\begin{figure}[t]
  \begin{center}
    \includegraphics[clip, width=8.4cm]{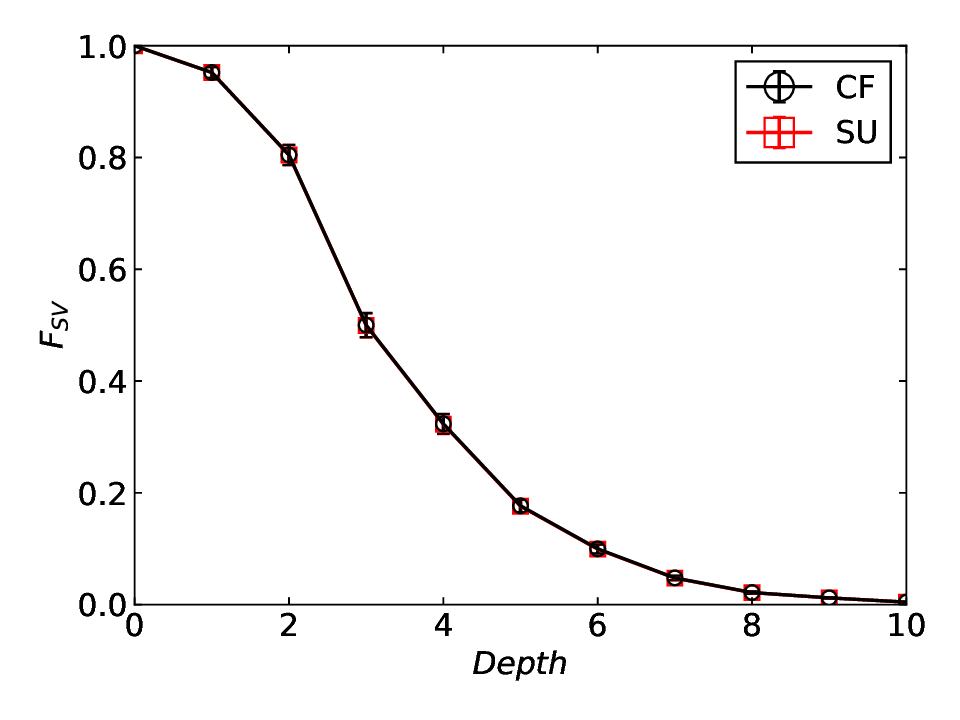}
    \caption{Numerically simulated fidelities $F_{\mathrm{SV}}$ for CF and SU.
             The fidelities for CF and SU are close.
             Note that CF refers to the CF-based tensor update.
             The error bars represent standard errors.
             }
    \label{fig:compare_fidelities}
  \end{center}
\end{figure}

Here, the approximation performance in quantum computer simulation is quantified by fidelity referenced to the quantum state given by SV-based quantum simulation as a performance metric:
\begin{align} \label{eq:fidelity_between_SV_and_TN}
F_{\mathrm{SV}} = |\langle\psi_{\mathrm{SV}}|\psi_{\mathrm{TN}}\rangle|^{2},
\end{align}
where ($|\psi_{\mathrm{SV}}\rangle$) and ($|\psi_{\mathrm{TN}}\rangle$) denote the qubit states given by SV-based quantum simulation and TN-based tensor update, respectively.
TN-based tensor update includes both SU and CF.
FIG.~\ref{fig:compare_fidelities} shows the average of $F_{SV}$ over $29$ random quantum circuits.
The performance difference between SU and CF is not significant.
As theoretically expected, in the region where $F_{\mathrm{CF}}$ was close to 1, $F_{\mathrm{SV}}$ was also high.
Moreover, even when $F_{\mathrm{CF}}$ is smaller than 1, there was no significant difference in $F_{\mathrm{SV}}$.
Further investigation is needed to determine whether $F_{\mathrm{SV}}$ remains close in general quantum circuits.

To compare the performance of SU and CF in various quantum circuits, we utilized QASM benchmark set \cite{QASM_bench, QASM_bench_paper}.
We used medium-scale quantum circuits because they have a sufficient number of qubits and allow for comparison with SV.
From the QASM benchmark set, quantum circuits that include mid-circuit measurements were excluded from the present evaluation.
The details of the quantum circuits used for the benchmark are provided in Table~\ref{tab:QASM_benchmark_set}.
To simulate the quantum circuits, two truncation parameters were employed: the number of singular values $\le 3$, and the singular values are truncated when the ratio relative to largest singular value is less than $0.0001$.

\begin{table}[h]
    \centering
    \caption{Quantum circuits used for benchmark. These quantum circuits are defined in QASM benchmark set. Refer to \cite{QASM_bench_paper} for detail information on these quantum circuits.}
    \label{tab:QASM_benchmark_set} 
    \begin{tabular}{ccc}
        \toprule
        Circuit Label & Benchmark & Number of qubits\\
        \midrule
        A & bigadder & 18 \\
        B & bv & 14 \\
        C & bv & 19 \\
        D & cat\_state & 22 \\
        E & cc & 12 \\
        F & dnn & 16 \\
        G & gcm & 13 \\
        H & ghz\_state & 23 \\
        I & ising & 26 \\
        J & knn & 25 \\
        K & multiplier & 15 \\
        L & multiply & 13 \\
        M & qec9xz & 17 \\
        N & qf21 & 15 \\
        O & qft & 18 \\
        P & qram & 20 \\
        Q & sat & 11 \\
        R & swap\_test & 25 \\
        S & wstate & 27 \\
        T & hhl & 14 \\
        U & factor247 & 15 \\
        V & vqe & 24 \\
        \bottomrule
    \end{tabular}
\end{table}

\begin{figure}[t]
  \begin{center}
    \includegraphics[clip, width=8.4cm]{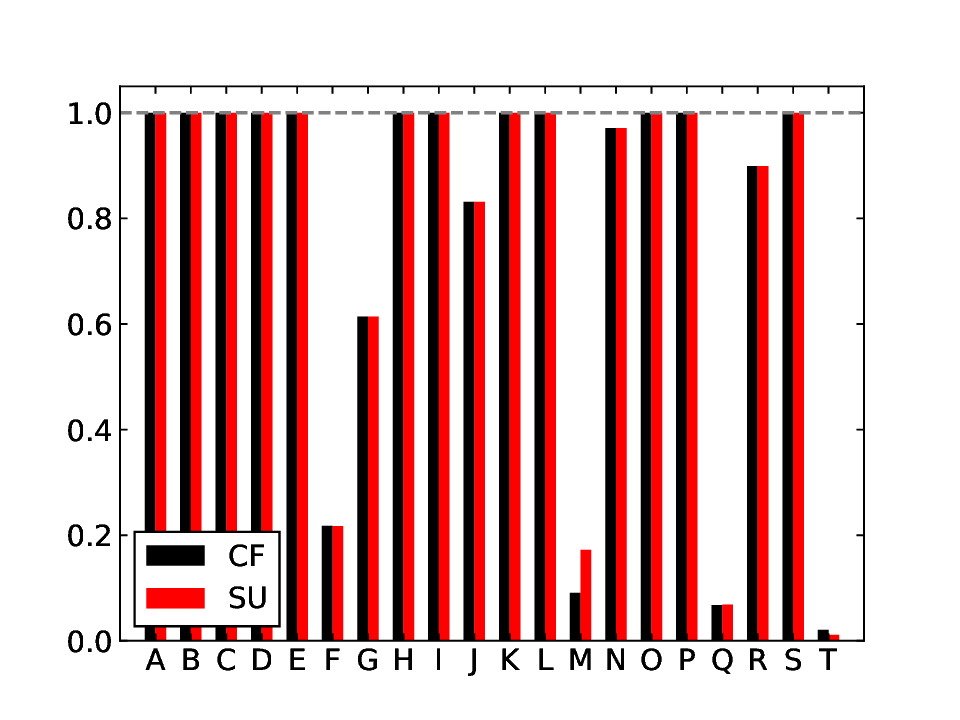}
    \caption{
             Numerically obtained fidelities $F_{\mathrm{SV}}$ given by CF and SU.
             The fidelities for CF and SU are close except for the quantum circuit M.
             Note that CF refers to the CF-based tensor update.
             }
    \label{fig:compare_fidelities_in_QASM_bench}
  \end{center}
\end{figure}

FIG.~\ref{fig:compare_fidelities_in_QASM_bench} shows the numerically obtained fidelities $F_{\mathrm{SV}}$ given by CF and SU.
The horizontal axis represents the circuit label shown in Table~\ref{tab:QASM_benchmark_set}.
In FIG.~\ref{fig:compare_fidelities_in_QASM_bench}, data with fidelities less than 0.0005 were excluded.
Quantum circuits U and V did not meet this fidelity threshold and are absent from FIG.~\ref{fig:compare_fidelities_in_QASM_bench}.
The fidelities $F_{SV}$ given by CF and SU are close except for the circuit M.
In the simulation of the circuit M, there is a difference between $F_{SV}$ given by SU and $F_{SV}$ given by CF.
On the other hand, except for the circuit M, CF and SU yield similar fidelities $F_{SV}$, which is consistent with the results shown in FIG.~\ref{fig:compare_fidelities}.

\begin{figure}[t]
  \begin{center}
    \includegraphics[clip, width=8.4cm]{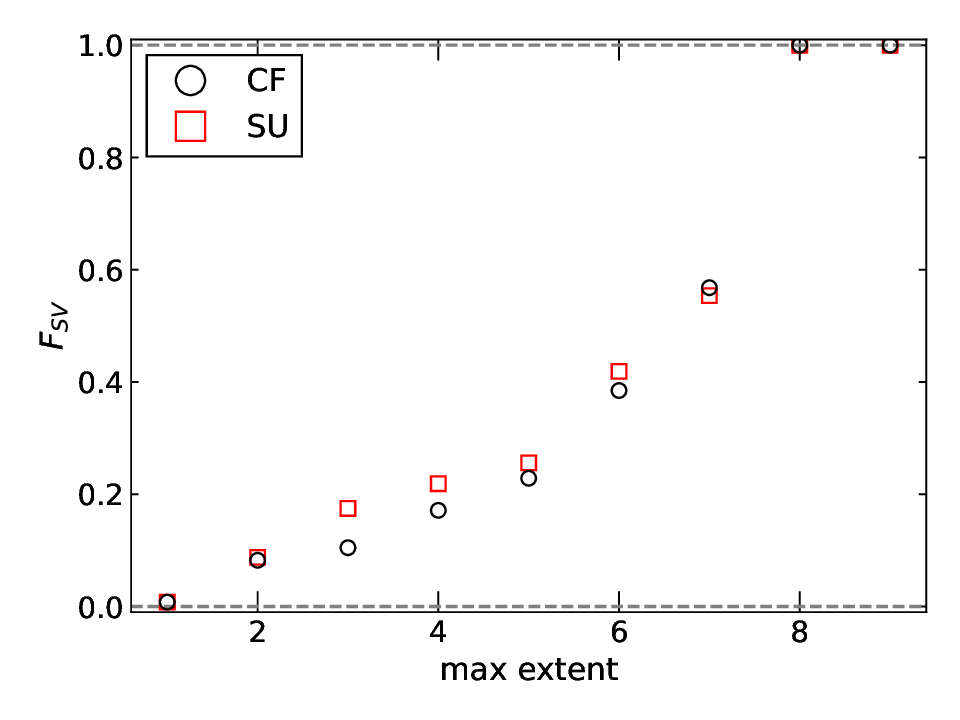}
    \caption{
             Numerically obtained fidelities $F_{\mathrm{SV}}$ given by CF and SU.
             There is no consistent superiority between CF and SU.
             Note that CF refers to the CF-based tensor update.
             }
    \label{fig:compare_fidelities_bond_dimensions_of_circuit_M}
  \end{center}
\end{figure}

To investigate details of the simulation of the circuit M, we change a truncation parameter.
FIG.~\ref{fig:compare_fidelities_bond_dimensions_of_circuit_M} shows the numerically obtained fidelities $F_{\mathrm{SV}}$ given by CF and SU in the simulation of the circuit M.
The horizontal axis represents maximum number of singular values.
There is no consistent superiority between CF and SU in the simulation of the circuit M.
This result implies that fidelities of CF and SU are influenced by factors other than fundamental performance.
We investigated the singular values in the simulation of the circuit M.
The singular values were found to be degenerate as shown in TABLE.~\ref{tab:degenerated_singular_values_of_M}.
Degenerate singular values might result in different states being selected in CF and SU during the truncation procedure, leading to variations in fidelity.
This investigation implies that even without a clear superiority in performance between CF and SU, accuracy could still vary.

\begin{table}[h]
    \centering
    \caption{
    A set of singular values generated by quantum gates in the circuit M, having 8 singular values.
    All singular values of this set are close to $1/\sqrt{8}$.
    This corresponds to a maximally entangled 3-qubit partition.
    }
    \label{tab:degenerated_singular_values_of_M}
    \begin{tabular}{ccc}
        \toprule
        singular value ID & 1st set & 2nd set\\
        \midrule
        0 & 0.35355339 \\
        1 & 0.35355339 \\
        2 & 0.35355339 \\
        3 & 0.35355339 \\
        4 & 0.35355339 \\
        5 & 0.35355339 \\
        6 & 0.35355339 \\
        7 & 0.35355339 \\
        \bottomrule
    \end{tabular}
\end{table}

\subsection{Discussion}\label{subsec:discussion}
Now we will discuss our results. 
Under less truncated conditions, the quantum states given by CF and SU are almost identical because SU maintains canonical conditions.
On the other hand, for various quantum circuits, the state given by SU may deviate from the state given by CF largely or fidelities may change depending on SU and CF.
However, our numerical experiments indicated that CF and SU yield similar fidelities $F_{SV}$.
Although there was one quantum circuit in which SU appeared to perform better, this may have been due to the degeneracy of singular values, and no intrinsic superiority or inferiority was found. 
In other words, SU exhibited no performance degradation relative to CF.

Next, let us discuss general quantum circuits.
It is difficult to simulate SV and $F_{\mathrm{SV}}$ for many qubits.
Moreover, behavior of $F_{\mathrm{CF}}$ is different from $F_{\mathrm{SV}}$ as shown in FIG.~\ref{fig:compare_fidelities_between_SU_and_CF} and FIG.~\ref{fig:compare_fidelities}, and $F_{\mathrm{CF}}$ can not assess superiority of CF and SU instead of $F_{\mathrm{SV}}$.
These limits prevent us from evaluating performance in large-scale quantum circuits.
Instead of large-scale quantum circuits, we employed quantum circuits that generates highly entangled states and medium-scale quantum circuits including various quantum gates.
In the results, no deterioration in performance of SU compared to CF was observed. 
SU is considered to be a promising method from the perspective of performance.

\section{CONCLUSION} \label{sec:conclusion}

In this paper, we investigated the simplification of MPS-based quantum computer simulation, especially focusing on its tensor update methods.
There are two aspects to be considered, i.e., performance (fidelity) and complexity (computational cost).
In quantum circuits with distant two-qubit gates, creating CF requires iterative operations of the QR decomposition, which causes high complexity.
Through the simulation for the quantum circuit with distant two-qubit gates between adjacent qubits, SU reduced the runtime by $O(N)$ compared to CF at almost identical fidelities.
More quantitatively, SU reduced the runtime by a factor of approximately $230$ compared to CF at fidelities around $1.0000$ in quantum circuits having $2000$ qubits and $2000$ two-qubit gates between adjacent qubits.
In addition, we evaluated performance under highly entangled conditions.
As theoretically expected, in the region where $F_{\mathrm{CF}}$ was close to 1, $F_{\mathrm{SV}}$ was also high.
Moreover, even when $F_{\mathrm{CF}}$ is smaller than 1, there was no significant difference in $F_{\mathrm{SV}}$.
To evaluate whether $F_{\mathrm{SV}}$ remains close in general quantum circuits, we tested various quantum circuits and computed fidelities referenced to the quantum state given by SV-based quantum simulation.
In most quantum circuits, the difference between the fidelities of SU and that of CF was not significant.
In a quantum circuit M, the fidelity of CF differs from that of SU.
However, the difference does not indicate superiority of either CF or SU. 
In other words, SU exhibited no performance degradation relative to CF.

We first discuss whether performance of SU relative to CF is close for general quantum circuits.
Our investigation, conducted on a variety of quantum circuits-including cases where the states given by SU and CF diverge, as well as circuits containing broad range of quantum gates-revealed that SU exhibited no performance degradation relative to CF.
These results suggest that SU is capable of delivering high-accuracy performance across a broad range of quantum circuits. 
Next, we discuss the types of quantum circuits for which SU is effective.
The high-accuracy performance of SU might imply that we can utilize SU instead of CF for general quantum circuits. 
However, SU does not reduce computational cost for general quantum circuits.
SU reduces computational cost of creating CF in quantum circuits with distant gates.
Instead of SU, optimization of quantum gate operation sequences is effective to reduce computational cost of creating CF.
However, the optimization process itself may present some challenges when dealing with large-scale and complex quantum circuits, and various optimization methods may be needed \cite{optimizing_tensors, optimizing_tensors_by_partition, optimizing_tensors_by_k_way}.
SU has potential to reduce complexity for the quantum circuits that contain many distant quantum gates and for which the optimization process is difficult.
Moreover, SU offers practical advantages: the computational time can be readily estimated from the number of gates, and the method is easy to parallelize, making it a suitable choice in many scenarios.

\begin{acknowledgments}
We would like to thank Tatsuya Sakashita, Atsushi Iwaki, Rihito Sakurai, and Hidemaro Suwa of Todo group in Tokyo University for providing academic insights.
We would like to express our gratitude to Isamu Kudo, Alifu Xiafukaiti, Yuki Takeuchi and Narumitsu Ikeda of Mitsubishi Electric Corp. for fruitful discussions.
We also wish to extend our thanks to George Monma, Tatsuo Kaniwa, Gaku Nagashima, and Yuichiro Minato from blueqat Inc. for providing the research environment and technical support.
This work was supported by the Center of Innovation for Sustainable Quantum AI (JST Grant Number JPMJPF2221).
\end{acknowledgments}

\bibliographystyle{apsrev4-2}
\bibliography{references}

\end{document}